\documentclass[prl,aps,floats,superscriptaddress,amsfonts,twocolumn,showpacs]{revtex4}

\usepackage{graphicx}

\begin{document}

\newcommand{\ket} [1] {\vert #1 \rangle}
\newcommand{\bra} [1] {\langle #1 \vert}
\newcommand{\braket}[2]{\langle #1 | #2 \rangle}
\newcommand{\proj}[1]{\ket{#1}\bra{#1}}
\newcommand{\mean}[1]{\langle #1 \rangle}
\newcommand{\opnorm}[1]{|\!|\!|#1|\!|\!|_2}

\title{Continuous-variable quantum key distribution protocols
over noisy channels}

\author{Ra\'{u}l Garc\'{\i}a-Patr\'{o}n}
\affiliation{QuIC, Ecole Polytechnique, CP 165,
Universit\'e Libre de Bruxelles, 1050 Bruxelles, Belgium}
\affiliation{Research Laboratory of Electronics, Massachusetts Institute
of Technology, Cambridge, MA 02139}

\author{Nicolas J. Cerf}
\affiliation{QuIC, Ecole Polytechnique, CP 165,
Universit\'e Libre de Bruxelles, 1050 Bruxelles, Belgium}

\begin{abstract}
A continuous-variable quantum key distribution protocol based on squeezed states
and heterodyne detection is introduced and shown to attain higher secret key rates
over a noisy line than any other one-way Gaussian protocol. This increased resistance
to channel noise can be understood as resulting from purposely adding noise to the signal
that is converted into the secret key. This notion of noise-enhanced tolerance to noise
also provides a better physical insight into the poorly understood discrepancies
between the previously defined families of Gaussian protocols.
\end{abstract}


\maketitle

Quantum Key Distribution (QKD) is a prominent application of
quantum information sciences, enabling two partners (Alice and Bob)
to share a secret key, which in turn allows them to communicate
with full security.
A particular class of QKD protocols based on the Gaussian modulation
of Gaussian states has attracted much attention over the last years
for its associated (homodyne or heterodyne) detection scheme offers
the prospect of very high key rates \cite{cerf-grangier}. In these
so-called continuous-variable (CV) protocols, the data which make the key
are encoded into continuous-spectrum quantum observables, namely
the quadrature components of a light field. These protocols fall
to date into three families, depending on which states
and detection schemes are used.

In the first one, Alice uses a source of {\em squeezed} states that are
randomly displaced along the squeezed quadrature, while Bob performs
{\em homodyne} detection \cite{Cerf01}. The experimental implementation
is much simplified with the second one, which is based at Alice's side
on {\em coherent} states modulated in both quadratures instead of squeezed states modulated along a single quadrature \cite{Gross02}. This second
proposal was first demonstrated experimentally in Ref.~\cite{Nature03},
while its implementation with optical telecom components was reported
in Ref.~\cite{Jerome05}.
In the third proposal, Alice still uses coherent
states but Bob performs {\em heterodyne} instead of homodyne detection,
measuring both quadratures together, hence eliminating the need
for an active random basis choice \cite{Ralph04}.

In this Letter, we introduce a fourth CV-QKD protocol based on {\em squeezed}
states and {\em heterodyne} detection, which, surprisingly,
happens to outperform all previous Gaussian protocols when the noise level
in the quantum channel is high.
This hitherto overlooked protocol, completing the family of Gaussian protocols,
had not been found earlier because, at first sight, it
serves no purpose measuring both quadratures when only one of them
(the squeezed quadrature) carries the key.
This striking effect can, however, be understood by exploring the analogy with
qubit-based QKD and realizing that adding some noise on the data
of the appropriate partner during the error correction
phase may result in an increase of the secret key rate~\cite{Renato05}.
We indeed can explain the improved resistance to noise of our new
protocol by using its equivalence with the first protocol,
based on squeezed states and homodyne measurement,
supplemented with noisy post-processing. This analysis also
allows us to construct the family of {\em optimal} Gaussian protocols
with respect to channel excess noise.

It has been known after Ref.~\cite{Renato05} that the performance
of qubit-based QKD protocols
can be increased by having Alice
adding some noise to her data in the error correction phase.
This additional classical noise makes the protocol more robust against noise
in the quantum channel because it is more detrimental to Eve than
to Bob. More precisely, for each quantum channel, there is
an optimal level of noise that Alice should add in order to
maximize the secret key rate.  An explanation of this phenomena can be found
in Ref.~\cite{Renes} using an entanglement-based description of BB84
together with a modified version of Shor-Preskill's unconditional
security proof \cite{Shor}.
Note that entangled-based CV-QKD protocols have also been introduced in Ref.~\cite{Silberhorn},
but entanglement will only be used in what follows
as a tool to analyze all four prepare-and-measure protocols on a same footing.
We shall show that this counterintuitive effect also appears, though
in disguise, in the case of CV-QKD protocols.
This is not straightforward, however, because of the distinction
that exists between direct reconciliation (DR) and reverse reconciliation (RR),
a feature which plays a central role in CV-QKD. In contrast to
qubit-based QKD, it is indeed crucial to specify whether Alice
or Bob is the reference during the error correction post-processing phase.
In DR, Alice plays this role and the maximal achievable range is known to be $3$dB \cite{Gross02};
in RR, there is no theoretical limitation to this range \cite{Nature03}.





In the following, we will focus on the security of CV-QKD
against collective attacks, where Eve interacts individually
with each signal pulse
sent by Alice but applies a joint measurement at the end of the classical post-processing stage. Studying this class of attacks is sufficient to prove unconditional security of qubit-based QKD protocols \cite{RenatoNature},
and we take for granted here the conjecture
that the same holds for CV-QKD \cite{OptGauss06}.
In addition, we restrict our study to Gaussian collective attacks
as they are known to be optimal \cite{OptGauss06}. Furthermore,
we consider RR as it works over longer distances. The corollary
is that Alice and Bob's roles must be interchanged when analyzing
the tolerance to noise (indeed, qubit-based QKD uses DR).
This leads us to introduce a fourth Gaussian protocol.



\begin{figure}[!t!]
\begin{center}
\includegraphics[width=9cm]{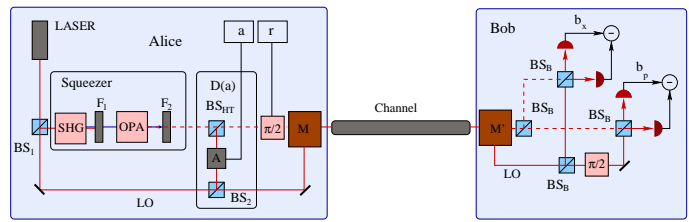}
\end{center}
\caption{Proposed experimental implementation of the new protocol.
The source (Alice) is based on a master laser beam.
A fraction of it is extracted to make the local oscillator (LO),
while the rest is converted into second harmonic in a nonlinear crystal (SHG). After spectral filtering (F$_1$), the second harmonic beam
pumps an optical parametric amplifier (OPA) which generates a squeezed vacuum state. Following the filtering of
the second harmonic (F$_2$), this squeezed state is displaced by $a$. This is done by mixing the state on a beamsplitter of high transmittivity ($T_{HT}\sim 99\%$ for $BS_{HT}$) with a coherent state
of intensity $a^2/(1-T_{HT})$, extracted from the LO.
The attenuation (A) thus depends on $a$, which is distributed
according to a Gaussian distribution $G(a)$ of variance $V_a$.
Before time multiplexing ($M$) the quantum signal
with the LO, Alice applies a phase shift $\theta=r\pi/2$ to it
depending on the value of the random bit $r$. Then, the two components of the time multiplexed signal travel to Bob through the same
fiber, thereby avoiding a spurious dephasing between the signal and LO.
At Bob's station, the two components are demultiplexed ($M'$), and
the quantum signal is heterodyne measured. The latter measurement consists in
splitting the quantum signal (and LO) in two with the balanced beamsplitters (BS$_B$), and then homodyning each beam.
The LO used in the second measurement
is dephased by $\pi/2$ in order to measure the conjugate quadrature.
Each homodyne detector is composed of a balanced beam splitter and a pair of highly efficient photodiodes; the difference of the photocurrents
gives the quadratures $b_x$ and $b_p$.}
\label{protocol4}
\end{figure}

\textit{The protocol.}
The first stage consists in quantum communication
over the quantum channel, characterized by the transmittivity $T$ and
added noise variance referred to the input $\chi_C$.
Alice generates a random bit $r$ and a real number $a$ drawn from
a Gaussian distribution $G(a)$ of variance $V_a$.
Subsequently, she generates a squeezed vacuum state of covariance matrix
${\rm diag}(1/V,V)$ and displaces it by an amount $(a,0)$,
see Fig.~\ref{protocol4}.
Before sending the state together with the local oscillator
through the quantum channel, she applies a random
dephasing of $\theta=r\pi/2$ to the state.
This dephasing is equivalent to randomly choosing to squeeze and displace either the $x$ or $p$ quadrature,
as in Ref.~\cite{Cerf01}. Averaging the output states over $G(a)$
gives the same (thermal) state for $r=0$ and $r=1$,
which prevents Eve from extracting information on which quadrature
was selected by Alice.
This imposes the constraint $V_a+1/V=V$ on Alice's modulation.
The quantum signal and local oscillator
can be transmitted over the same fiber by using
a time multiplexing technique, as in Ref.~\cite{Jerome07}.
At Bob's station, the signal is first demultiplexed and subsequently measured
by a standard heterodyne measurement, as shown in Fig.~\ref{protocol4}.
The use of heterodyning makes the random number generator
on Bob's side unnecessary since there is no need to switch between
the measurements of conjugated bases, just as in Ref.~\cite{Ralph04}.
After repeating these steps many times, Alice ends up with
a long string of data $a$ correlated with Bob's heterodyne
data $(b_x,b_p)$.

The second stage is the classical post-processing stage, which
serves extracting the secret key. It
starts by Alice revealing the string of random bits $r$ encoding her
chosen quadratures and Bob keeping as his final string of data $b$
the measurements ($b_x$ or $b_p$) matching Alice's choices.
This step is followed by the channel estimation,
where Alice and Bob reveal a fraction of their data in order to estimate
$T$ and $\chi_C$,
which allows them to bound Eve's information.
Subsequently, Alice and Bob apply a RR algorithm, such as LDPC codes \cite{Jerome07} combined with a discretization operation. This operation outputs two perfectly correlated binary strings.
Finally, both partners apply a privacy amplification algorithm based, e.g.,
on hash functions \cite{Jerome07}, which produces a shared binary secret key
from their perfectly correlated data.
As shown in Ref.~\cite{Renato05}, the achievable RR secret key rate reads
\begin{equation}
K=I(a{\rm:}b)-S(b{\rm:}E),
\label{Kcoll}
\end{equation}
where $I(a{\rm:}b)$ is the Shannon information between Alice
and Bob's data while $S(b{\rm:}E)$ is Eve's information on $b$
given by the Holevo quantity $S(b{\rm:}E)=S(\rho_E)-\int db\; p(b)S(\rho_E^b)$.
Given that Eve can be assumed to hold the purification of the system
and that Gaussian attacks are optimal, we can directly compute $K$ from
the covariance matrix $\gamma_{AB}$ inferred from
the channel estimation.

\begin{figure}[!t!]
\begin{center}
\includegraphics[width=8.5cm]{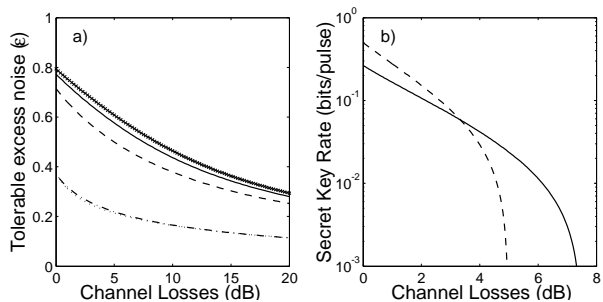}
\end{center}
\caption{a) Tolerable excess noise $\epsilon$ (in shot-noise units)
as a function of the channel losses (in dB) for RR protocols:
new (solid line), squeezed states and homodyning (dashed line) \cite{Cerf01},
coherent states and homodyning (dotted line) \cite{Gross02},
coherent states and heterodyning protocol (dash-dotted line) \cite{Ralph04}.
The optimal protocol with Bob's added noise $\chi_D$ is also shown (crosses).
The curves are plotted for $V\rightarrow\infty$. b) Secret key rates as a function of the channel losses (in dB)
for RR protocols: new (solid line), squeezed states and homodyning
protocol (dashed line) \cite{Cerf01}. The curves are plotted for an excess noise $\epsilon=0.5$ and $V=40$.}
\label{NewFigure}
\end{figure}

\textit{Tolerance to noise.}
In Fig.~\ref{NewFigure} a), we show that this new protocol performs better
than all previous RR protocols in term of tolerable excess noise, i.e., the lowest $\epsilon=\chi_C-(1-T)/T$
that gives a zero secret key rate.
In realistic implementations of CV-QKD, the excess noise generally comes from
the laser's phase noise and imperfections in the modulation,
as discussed in Ref.~\cite{Jerome05}, so that it can be considered
as approximately independent of the length of the fiber.
This does not mean, however, that the new protocol gives higher rates
regardless of the channel transmittivity. As shown in Fig.~\ref{NewFigure} b),
it is only for losses higher than a given threshold  that it
gives a higher secret key rate than the protocol of Ref.~\cite{Cerf01}.

In the new protocol, Bob disregards either $b_x$ or $b_p$ during the
post-processing stage, depending on Alice's quadrature choice $r$.
This is equivalent to tracing out the mode that is not used
in Bob's heterodyne measurement, so that the new protocol can be viewed
as a noisy version of the protocol based on squeezed states and homodyne measurement \cite{Cerf01} where Bob  inserts a balanced beamsplitter
before his measurement.
The losses induced by this beamsplitter translate into noise once Bob classically amplifies his outcome to match the initial signal.
Therefore, we have a clear demonstration that adding noise that is
not controlled by Eve on Bob's side can be beneficial in CV-QKD
for a RR protocol.

\begin{figure}[!t!]
\begin{center}
\includegraphics[width=8cm]{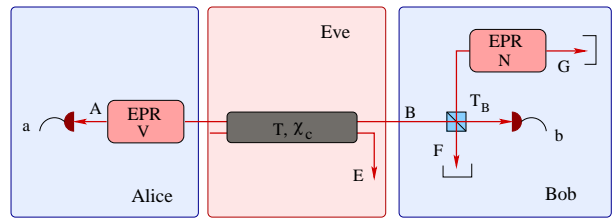}
\end{center}
\caption{Entanglement-based description of the protocol
with general Gaussian added noise on Bob's side.
The source of squeezed states on Alice's side is replaced by
an entangled pair (EPR) of variance $V$,
followed by an homodyne measurement of mode $A$. The other mode is sent to Bob
through the quantum channel. Before Bob's homodyne detection,
the state received by Bob is mixed with a thermal state (half of an EPR pair)
of variance $N$ on a beamsplitter of transmittivity $T_B$($\chi_D=(1-T_B)N/T_B$).}
\label{scheme2}
\end{figure}

Interestingly, this effect has a counterpart in DR which remained unnoticed
to date although it is visible with known protocols. We indeed observe
in Ref.~\cite{Navascues05} that the homodyne protocol
based on coherent states gives a better tolerance to excess noise in DR than
that based on squeezed states. The reason is that the former protocol \cite{Gross02} can actually be viewed as a noisy version of the latter
protocol \cite{Cerf01}, where the noise is now added by the same mechanism
but on Alice's side in the entanglement-based equivalent scheme, see Fig.~\ref{scheme2}.
In this scheme, coherent states are prepared by
Alice applying an heterodyne measurement, which can be viewed as
a noisy homodyne measurement.
We thus conclude that there is a beneficial
effect of noise if added on the reference side of error correction
(Alice in DR and Bob in RR). Clearly, adding noise on the other side
is always detrimental as it decreases the information
between the authorized parties without
affecting the eavesdropper's information.

\begin{figure}[!t!]
\begin{center}
\includegraphics[width=8.5cm]{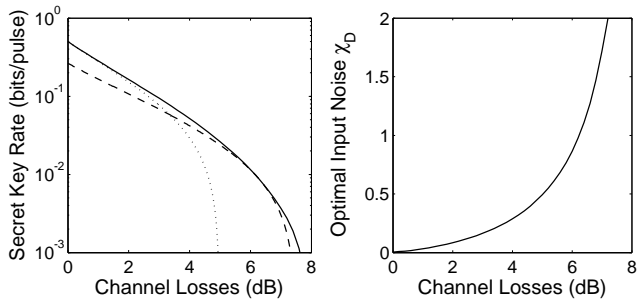}
\end{center}
\caption{a) Optimal secret key rates as a function of the channel losses (dB)
for a fixed excess noise $\epsilon=0.5$ (solid line) compared to the
protocol based on squeezed states and homodyning \cite{Cerf01} (dotted line)
and the new protocol proposed in this Letter (dashed line).
b) Optimal choice of $\chi_{D}$ (in shot-noise units) that maximizes the secret key rate. The curves are plotted for $V=40$.}
\label{OptimalRate}
\end{figure}

\textit{Optimal protocol.}
We now generalize the above new RR protocol to optimally resist
against an arbitrary channel noise.
In Figure~\ref{scheme2}, we exhibit an entanglement-based description
of CV-QKD protocols, where Bob replaces his heterodyne measurement
by an ideal homodyne measurement preceeded by
a general Gaussian phase-insensitive added noise.
This models the following physical situations:
i) inefficient homodyne detection with efficiency $T_B$ and electronic noise variance $v=(1-T_B)(N-1)$;
ii) perfect homodyne detection followed by a classical Gaussian added
noise of variance $\chi_D=(1-T_B)N/T_B$;
iii) any combination of the previous cases giving the same $\chi_D$.
The secret key rate
can be calculated using the following technique.
First we use the fact that Eve's system $E$ purifies $AB$, that is,
$S(E)=S(AB)$.
Secondly, after Bob's projective measurement yielding $b$,
the system $AEFG$ being pure, we have
$S(E|b)=S(AFG|b)$. For Gaussian states
$S(AFG|b)$ is the same for all $b$'s, being just a function of the covariance matrix
$\gamma_{AB}$. Thus, we obtain,
\begin{equation}
K=I(a{\rm:}b)-S(AB)+S(AFG|b)
\end{equation}
which can be calculated from the covariance matrix
\begin{equation}
\gamma_{AB}=\left [\begin{array}{cc}
x\mathbb{I}    & z \sigma  \\
z \sigma & y \mathbb{I}      \\
\end{array}\right ],
\end{equation}
where $x=V$, $y=T(V+\chi_C)$, $z=\sqrt{T(V^2-1)}$,
$\mathbb{I}={\rm diag}(1,1)$ and $\sigma={\rm diag}(1,-1)$.
The information between Alice and Bob reads
\begin{equation}
I(a{\rm:}b)=\frac{1}{2}\log\Bigg[\frac{V+\chi}{\chi+1/V}\Bigg],
\end{equation}
where $\chi=\chi_C+\chi_D/T$.
Then, $S(AB)$ is a function of the symplectic eigenvalues $\lambda_{1,2}$ of $\gamma_{AB}$ which reads
\begin{equation}
S(AB)=G\big[(\lambda_1 -1)/2\big]+G\big[(\lambda_2 -1)/2\big],
\label{S}
\end{equation}
where $G(x)=(x+1)\log (x+1)-x\log x$ is the von Neumann entropy of a thermal state and
\begin{equation}
\lambda_{1,2}^{2}=\frac{1}{2}\bigg[\Delta\pm \sqrt{\Delta^2-4D^2}\bigg].
\end{equation}
Here, we have used the notation $\Delta=x^2+y^2-2z^2$ and $D=xy-z^2$.
Finally, $S(AFG|b)$ is a function of the symplectic eigenvalues $\lambda_{3,4}$ similar to (\ref{S})
where $\lambda_{3,4}^2$ are solutions of the second order polynomial $\lambda^4-A\lambda^2+B=0$ with
\begin{eqnarray}
A&=&\frac{1}{y+\chi_D}\big[y+xD+\chi_D\Delta\big] \\
B&=&\frac{D}{y+\chi_D}\big[x+\chi_DD\big].
\end{eqnarray}
By tuning Bob's added noise $\chi_D$, it is possible to maximize
the secret key rate, as shown in Fig.~\ref{OptimalRate}. More importantly,
the resulting family of optimal protocols exhibits the highest tolerance
to noise among all Gaussian CV-QKD protocols, as demonstrated in
Fig.~\ref{NewFigure} a) (crosses).

Note that although Bob's heterodyne measurement is useful
to get an insight on this enhanced tolerance to noise, Bob eventually
disregards one of the two quadratures in the actual protocol. Thus,
up to a factor of two in the key rate, he may as well apply a (noisy)
homodyne measurement and keep the outcome only when he has measured
the right quadrature. Finally, instead of using a random numbers generator
to generate the noise $\chi_D$, it is certainly more interesting for Bob
to  do it physically by tuning the efficiency of his detector.


\textit{Conclusion.}
We have proposed a new CV-QKD protocol using squeezed states and
heterodyne detection, which outperforms all known Gaussian protocols
in terms of resistance to noise. This enhanced robustness can be
interpreted as the continuous-variable counterpart of the effect,
first described in  Ref.~\cite{Renato05},
that adding noise in the error-correction post-processing phase
may increase the secret key rate of one-way qubit-based protocols.
Then, we have studied the impact of a general Gaussian phase-insensitive
noise on the secret key rate, and have shown that for each quantum channel (characterized by its transmittivity $T$ and added noise variance $\chi_C$), there is an optimal noise $\chi_D$ that Bob must add
to maximize the secret key rate. The resulting protocol also exhibits
the highest resistance to noise among all Gaussian protocols.
This noise-enhanced tolerance to noise is particularly interesting
for reverse-reconciliation CV-QKD protocols, which work over larger distances,
but, interestingly, it also has an analogue for direct-reconciliation protocols. This gives a physical explanation to the previously observed -- but poorly understood -- discrepancies between the efficiencies of Gaussian protocols.


We acknowledge financial support from the EU under projects
COMPAS and SECOQC,
from the Belgian foundation FRIA,
and from the W. M. Keck Foundation Center for Extreme Quantum Information Theory.

\end{document}